\documentclass[a4paper,12pt]{article}

\usepackage{pstricks}
\usepackage{amsmath}\usepackage{amssymb, graphics}
\usepackage{latexsym}\usepackage{cite}
\usepackage{amsmath}\usepackage{amssymb}
\usepackage{latexsym}
\usepackage[dvips]{graphicx}
\usepackage{pstricks}
\usepackage{bm}
\usepackage{axodraw}
\newcommand{\be}{\begin{equation}}
\newcommand{\ee}{\end{equation}}

\def\beq{\begin{equation}}
\def\eeq{\end{equation}}

\def\al{\alpha}
\def\bt{\beta}
\def\Ga{\Gamma}

\def\de{\delta}
\def\De{\Delta}

\def\te{\theta}

\def\lam{\lambda}

\def\om{\omega}
\def\ep{\epsilon}

\def\sq{\sqrt}

\def\l{\left (}
\def\r{\right )}

\def\fr{\frac}
\def\la{\label}
\def\hs{\hspace}
\def\vs{\vspace}

\def\ran{\rangle}
\def\lan{\langle}

\def\tl{\tilde}

\begin{document}
\baselineskip=18pt

\begin{flushright}
OSU-HEP-09-01\\
January 8, 2009
\end{flushright}

\vs{0.7cm}

\begin{center}
{\Large\bf
New Ways to Leptogenesis with Gauged $B-L$ Symmetry}
\end{center}

\vspace{0.5cm}
\begin{center}
{\large
{}~K.S. Babu\footnote{E-mail: babu@okstate.edu},{}~
Yanzhi Meng\footnote{E-mail: yanzhi.meng@okstate.edu},
{}~Zurab Tavartkiladze\footnote{E-mail: zurab.tavartkiladze@okstate.edu}
}
\vspace{0.5cm}

{\em Department of Physics, Oklahoma State University, Stillwater, OK 74078, USA }
\end{center}
\vspace{1.6cm}

\begin{abstract}
We show that in supersymmetric models with gauged $B-L$ symmetry, there is a new source
for cosmological lepton asymmetry.  The Higgs bosons responsible for
$B-L$ gauge symmetry breaking decay dominantly into right--handed sneutrinos $\tilde{N}$
and $\tilde{N}^*$ producing an asymmetry in $\tilde{N}$ over $\tilde{N}^*$.  This can be fully  converted into
ordinary lepton asymmetry in the decays of $\tilde{N}$.  In simple models with gauged $B-L$ symmetry
we show that resonant/soft leptogenesis is naturally realized.  Supersymmetry
guarantees quasi--degenerate scalar states, while soft breaking of SUSY provides the needed
CP violation.  Acceptable values of baryon asymmetry are obtained without causing serious problems
with gravitino abundance.
\end{abstract}

\vs{0.7cm}


\newpage

\section{Introduction}

Baryon number minus lepton number ($B-L$) is a non--anomalous symmetry in the standard model.
There is a perception that all non--anomalous symmetries may have a gauge origin. $B-L$ may then
be a true gauge symmetry broken spontaneously at a high energy scale.  Such a scenario fits
well with the small neutrino masses observed in experiments.  This is because gauging of $B-L$
requires the introduction of right--handed neutrinos $N_i$, one per family, for canceling the triangle anomaly
associated with $[U(1)_{B-L}]^3$.  These $N_i$ fields facilitate the seesaw mechanism \cite{seesaw}
to generate small neutrino masses.  In this context one is able to relate the mass of the heavy right--handed
neutrino to the scale of $B-L$ symmetry breaking.  With just the standard model gauge symmetry
the right--handed neutrinos are not compelling, and even if they are introduced, their bare Majorana masses are
not protected and can take values as large as the Planck mass.

In the supersymmetric context there is yet another motivation for gauging $B-L$.  It would lead to
a natural understanding of $R$--parity \cite{mohap,goran}.  This can be seen by writing the $R$--parity transformation
as $R = (-1)^{3(B-L)+2S}$, which clearly shows the close relation between $R$ parity and $B-L$.
If the $B-L$ gauge symmetry is broken by Higgs fields carrying even number of $B-L$ charge, then a discrete
$Z_2$ symmetry will remain unbroken, which would serve as $R$--parity. Such Higgs fields are just
the ones needed for generating large Majorana neutrino masses for the right--handed
neutrinos, which requires $B-L$ breaking by two units.  $R$--parity is usually assumed in MSSM as an ad hoc
symmetry, in order to avoid rapid proton decay and to identify the lightest SUSY particle as the cosmological dark matter.  These are natural consequences of gauged $B-L$ symmetry.  This symmetry
also fits inside of $SO(10)$ grand unification, which is very well motivated because of the unification
of quarks and leptons of a family into a single multiplet.  It is well known that with or without
supersymmetry, existence of right--handed neutrinos can explain the observed excess of baryons  over
antibaryons in the universe via leptogenesis \cite{FY}. The $N$ field decays into leptons, generating
an asymmetry in lepton number, which is converted to baryon asymmetry by electroweak
sphalerons \cite{kuzmin}.  (For reviews on  leptogenesis see \cite{giudice,buchmuller}.)

In this paper we investigate baryogenesis via leptogenesis in supersymmetric models with gauged $B-L$ symmetry.
We have identified a new source for leptogenesis in this context.  The Higgs fields that
spontaneously break $B-L$ symmetry produce an excess of $\tilde{N}$ over $\tilde{N}^*$ in their decays, where
$\tilde{N}$ stands for the scalar partner of the right-handed neutrino $N$.  This asymmetry in
$\tilde{N}$ is converted into ordinary lepton asymmetry when the $\tilde{N}$ decays into leptons
and Higgs bosons.  The electroweak sphalerons convert this lepton asymmetry into baryon asymmetry.

In this scenario, one realizes resonant \cite{resonant1,resonant2,resonant3} and soft leptogenesis \cite{soft1,soft2}.
Resonant leptogenesis assumes
nearly degenerate states (fermions or scalars) that decay into leptons producing an asymmetry which is resonantly enhanced.  
Usually the needed degeneracy is achieved by postulating additional symmetries.  In our context, supersymmetry
guarantees near degeneracy of the Higgs states.  This comes about since in the SUSY limit, the Higgs
scalars responsible for $B-L$ symmetry breaking form partners of a Dirac fermion, leading to two complex (or
four real) degenerate scalar states.  Once SUSY breaking is turned on, this degeneracy is lifted, but by terms
that are suppressed by a factor $M_{\rm susy}/M_\Delta$, where $M_\Delta$ denotes the mass of the
decaying heavy Higgs particle.  In the simplest model with gauged $B-L$ symmetry, CP violation needed for leptogenesis
is provided by soft SUSY breaking effects.  Thus the model realizes soft leptogenesis.  We compute the baryon
asymmetry generated through this $\tilde{N}$ asymmetry in a simple model with gauged $B-L$ symmetry.
As in soft leptogeneis, we find that for a range of soft SUSY breaking parameters, reasonable values
of baryon asymmetry can be generated.  This mechanism works well when the mass of the decaying
Higgs filed is less than about $10^8$ GeV.  The Davidson--Ibarra bound \cite{davidson}, which requires the decaying
right--handed neutrino to be heavier than $10^9$ GeV in conventional leptogeneis, is evaded
in our framework because the source of CP violation resides in SUSY breaking couplings.  Such a bound
causes a problem with gravitino abundance \cite{gravitino1,gravitino2}, which requires the reheat temperature after inflation
to be $T_R < 10^7$ GeV. Our scenario does not have the gravitino problem, since the mass of the heavy
Higgs particle is $< 10^8$ GeV.  Some of
the soft SUSY parameters have to take unusually small values, a situation common with soft
leptogenesis, although the parameters that are small in our models are different ones, associated
with $B-L$ symmetry breaking.

We present the minimal gauged SUSY model in Sec. 2, work out the spectrum of the model after SUSY
breaking in Sec. 3, and compute the cosmological lepton asymmetry in Sec. 4.

\section{Minimal Supersymmetric Gauged $B-L$ Model}

The minimal supersymmetric model with gauged $B-L$ symmetry extends  the gauge group of
MSSM to  $SU(3)_C \times
SU(2)_L \times U(1)_Y \times U(1)_{B-L}$.  The triangle anomaly associated with $[U(1)_{B-L}]^3$ is canceled
by contributions from right--handed neutrinos $N_i$, which must exist, one per family.  Since the $N_i$ fields should
be much heavier than the weak scale in order for the seesaw mechanism for small neutrino masses  to be effective,
we assume that $B-L$ symmetry is broken in the SUSY limit.  The simplest set of scalar superfields that would achieve
this -- if one insists, as we do, on renormalizable coulings -- is $\{\overline{\Delta},~\Delta,~S\}$, where the first two fields carry $B-L$ charges of $\pm 2$, while
$S$ is neutral.   All three fields are neutral under $SU(3)_C \times SU(2)_L \times U(1)_Y$.
The $B-L$ charge of the $\Delta$ field is chosen so that it has direct Yukawa couplings with the $N$ fields, which would provide large Majorana masses for them upon spontaneous symmetry breaking. This choice also guarantees that $R$--parity
of MSSM will remain unbroken even after spontaneous symmetry breaking, since $\left\langle \Delta \right
\rangle \neq 0$ leaves an unbroken $Z_2$ symmetry, which functions as $R$--parity.
Our normalization of $B-L$ charge is as follows.  $(N, ~e^c)$ have charge $+1$, $L$ has
charge $-1$, $Q$ has charge $1/3$ while $(u^c,~d^c)$ fields carry charge $-1/3$.  No other fields beyond MSSM fields are introduced.

The superpotential of the model consistent with the extended gauge symmetry is given by
\begin{eqnarray}
W &=& W_{\rm MSSM}+W^{(B-L)}~, \nonumber \\
W^{(B-L)}&=&\lam S(\De \bar{\De }-M^2)+\fr{1}{2}f_{ij}N_iN_j\De +Y_{\nu }^{\al i}L_{\al }N_iH_u~.
\la{totW}
\end{eqnarray}
Here $W_{\rm MSSM}$ is the MSSM superpotential. $L_{\al}$ denotes the left--handed lepton doublets,
$H_u$ is the up--type Higgs doublet, and $i,\alpha$ are family indices.
Note that all $R$--parity violating couplings are forbidden in the
superpotential by the $B-L$ symmetry.  The Majorana masses for the right--handed neutrinos arise only after spontaneous
breaking of $B-L$ symmetry after $\left\langle \Delta \right \rangle \neq 0$ develops, via the
couplings $f_{ij}$.  The Dirac Yukawa
couplings $Y_{\nu }$ will then generate small neutrino masses via the seesaw mechanism.
Bare mass terms for $S$ as well as for $\Delta \overline{\Delta}$ and an $S^3$ term
have not been written in Eq. (\ref{totW}).  This is for simplicity and their omission can be justified
by invoking an $R$ symmetry.

We minimize the potential, which contains $F$--terms resulting from Eq. (\ref{totW}) and a $D$--term
corresponding to the $B-L$ symmetry, in the SUSY limit.  Demanding the vanishing of
$F$--terms, $F_S=F_{\De }=F_{\bar{\De }}=0$, yields  $\lan S\ran =0$ and $\lan \De \bar{\De }\ran =M^2$. The
vanishing of the $D$--term implies $|\De |=|\bar{\De }|$.
Without loss of generality we choose $\lan \De \ran =|M|$.  Consequently we have $\lan \bar{\De } \ran =|M|e^{i\phi_{M^2}}$, with the definition $\phi_{M^2} \equiv {\rm arg}(M^2)$.  The spectrum of the
model in the SUSY limit consists of a massive vector multiplet ${\cal V}_B$ and a pair of degenerate
chiral multiplets $(\Delta_0,~S)$ with masses given by
\begin{equation}
M_{\cal V_B} = 2 g_B |M|,~~~~~~~~M_\Delta = \sqrt{2}|\lambda| |M|~.
\label{mass}
\end{equation}
Here $g_B$ denotes the $B-L$ gauge coupling.  In this limit, the $B-L$ gaugino pairs up with
a Higgsino (denoted $\Delta'$) which is a linear combination of $\Delta$ and $\overline{\Delta}$
fields.  The orthogonal combination $\Delta_0$ pairs up with the $S$--Higgsino to forma a
Dirac fermion.  Small SUSY breaking effects, to be discussed shortly, will split the masses of
the two Weyl components in each of these Dirac fermions.
The ($\Delta_0,~S)$ system consists of two complex scalars as well -- corresponding
to four real nearly degenerate scalar states once small SUSY breaking effects are included, which
are physical.  It is these nearly degenerate scalar states that will be relevant for leptogenesis.

We will be interested in the limit where the physical Higgs multiplet $(\Delta_0,~S)$ is somewhat
lighter than the gauge supermultiplet, that is, in the limit $\sqrt{2}\lambda \ll 2 g_B$.  Precisely
how much lighter will be quantified later, but we will not need a larger hierarchy in masses,
$M_\Delta < 0.1 ~M_{\cal V_B}$ or so will suffice.  With such a mild hierarchy in masses, the dominant
decay of the $(\Delta,~S)$ Higgs fields will be into right-handed neutrino fields.  This will enable
a new way of generating lepton asymmetry stored in $\tilde{N}$ fields.  With $M_\Delta \ll  M_{\cal V_B}$,
we can integrate out the vector supermultiplet to obtain an effective superpotential $W_{\rm eff}$ and
an effective K\"{a}hler potential $K_{\rm eff}$ involving only the $(\Delta_0,~S)$ fields and the MSSM
superfields.

To obtain the effective Lagrangian of the theory after integrating out the vector superfield, we work in the unitary gauge and make supersymmetric transformations
on the ($\Delta,~\overline{\Delta})$ fields, the gauge vector multiplet ${\cal V}_B$, and all fields $\Phi_i$
carrying $B-L$ charge $q_i$ to go to a new basis with $(\Delta',~\Delta_0)$ fields and a shifted ${\cal V}_B$ gauge superfield:
\begin{eqnarray}
\De &=&(|M|+\fr{1}{\sq{2}}\De_0)e^{q_{\De }g_B\De'}~,~~~~~~~\bar{\De }=(|M|+\fr{1}{\sq{2}}\De_0)e^{-q_{\De }g_B\De' +i\phi_{M^2}}~,~~~ \nonumber \\
{\cal V}_B &=& {\cal V}_B^0-\De'-{\De'}^{\dag }~,~~~~~~~~~~~~~~~~~\Phi_i \to e^{q_ig_B\De'}\Phi_i ~.
\la{subs}
\end{eqnarray}
We have kept the $B-L$ charge
of $\Delta,~\overline{\Delta}$ fields as $(q_\Delta,~-q_\Delta)$ to be more general.

With these redefinitions, the original K\"{a}hler Lagrangian, given by
\beq
{\cal L}_D^{(B-L)}=\int  d^4\te \l \De^{\dag }e^{q_{\De }g_B{\cal V}_B}\De +\bar{\De }^{\dag }e^{-q_{\De }g_B{\cal V}_B}\bar{\De }+
\sum_i\Phi_i^{\dag }e^{q_ig_B{\cal V}_B}\Phi_i\r ~
\la{L-D}
\eeq
transforms into
\beq
{\cal L}_D^{(B-L)}=\int  d^4\te \l \left\{2|M|^2+\sq{2}|M|(\De_0+\De_0^{\dag })+\De_0^{\dag }\De_0\right\}{\rm Cosh} (q_{\De }g_B{\cal V}_B^0) +
\sum_i\Phi_i^{\dag }e^{q_ig_B{\cal V}_B^0}\Phi_i\r ~.
\la{L-D1}
\eeq
Observe that the $\De'$ field has disappeared in Eq. (\ref{L-D1}), it has been eaten up by the gauge superfield ${\cal V}_B^0$. In the process the gauge field ${\cal V}_B^0$ becomes massive, all its components acquiring
a mass $M_{\cal V_B}^2=q_{\De }^2g^2|M|^2$, as can be readily seen by expanding the Cosh function in Eq. (\ref{L-D1}).

Now we can integrate out the massive gauge superfiled ${\cal V}_B^0$. We obtain the following
effective K\"{a}hler Lagrangian:
\begin{eqnarray}
{\cal L}_{\rm D, eff}^{(B-L)}&=&\int  d^4\te \left [ \De_0^{\dag }\De_0+\sum_i\Phi_i^{\dag }\Phi_i-
\fr{1}{4q_{\De}^2|M|^2}\l \sum_iq_i\Phi_i^{\dag }\Phi_i\r^2\right. \nonumber \\
&~&\left. +\fr{\De_0\!+\!\De_0^{\dag }}{4\sq{2}q_{\De}^2|M|^3}\l \!\sum_iq_i\Phi_i^{\dag }\Phi_i\!\r^2\!\!\!+\hs{-0.1cm}
\fr{1}{8q_{\De}^2|M|^4}(\De_0^{\dag }\De_0 \!-\De_0^2\!-\De_0^{\dag 2})\l \!\sum_iq_i\Phi_i^{\dag }\Phi_i\!\r^2 \hs{-0.2cm}+\cdots \right ],
\label{Kahler}
\end{eqnarray}
where the $\cdots$ indicate terms with higher powers of $1/|M|$. Eq. (\ref{Kahler}) describes the interactions
of the light $\Delta_0$ field with other light MSSM fields through the exchange of the gauge supermultiplet.  Notice that these interactions are suppressed by $1/|M|^3$.

With the redefinition of fields given in Eq. (\ref{subs}), the superpotential of Eq. (\ref{totW})
becomes $W_{\rm eff}=W_{\rm MSSM}+W(\De_0,N)$ with
\beq
W(\De_0,N)=\lam S e^{i\phi_{M^2}}\l |M|\sq{2}\De_0+\fr{1}{2}\De_0^2\r +\fr{1}{2}f_{ij}(|M|+\fr{1}{\sq{2}}\De_0)N_iN_j+Y_{\nu }^{\al i}L_{\al }N_iH_u~.
\la{Wef}
\eeq
Note that the $\Delta'$ field has disappeared in Eq. (\ref{Wef}).  Majorana masses for $N$ have been generated
with $M_{N_i} = |f_i||M|$, where $|f_i| $ are the real and diagonal eigenvalues of the matrix $f_{ij}$.  It is also
clear from Eq. (\ref{Wef}) that $(\tilde{\Delta}_0,~\tilde{S})$ fields pair up to form a Dirac fermion with a mass given by $M_\Delta = \sqrt{2}|\lambda| |M|$.  Their scalar partners $(\Delta_0,~S)$ are of course degenerate with
these fermions, since SUSY breaking has not yet been turned on.

We assume that at least one of the $N_i$ fields is lighter than $\Delta_0$.  Such situation is quite natural,
especially when the $N_i$ fields have hierarchical masses.  We denote this light $N_i$ field simply as
$N$ (assuming for simplicity that only one such field is lighter than $\Delta_0$) with its mass given by $M_N = |fM|$.  The dominant decays of $\Delta_0$ scalar will then be $\Delta_0 \rightarrow \tilde{N}+\tilde{N}$,
$\Delta_0 \rightarrow \tilde{N}^*\tilde{N}^*$, and $\Delta_0 \rightarrow NN$.  There is also a subdominant
decay of $\Delta_0$ into $\tilde{N}\tilde{N}^*$. Here $N$ denotes the right--handed
neutrino, while $\tilde{N}$ stands for its scalar partner.  Supersymmetry will dictate that the decays of the fermionic partner of $\Delta_0$, denoted as $\tilde{\Delta}_0$ will be to $\tilde{N} N$ and $\tilde{N}^* N$ final states with an identical width.  The total width
for the decays of the scalar $\Delta_0$ is given by
\begin{equation}
\Gamma(\Delta_0 \rightarrow \tilde{N}\tilde{N}+ \tilde{N}^*\tilde{N}^*+\tilde{N} \tilde{N}^*+NN) = {|f|^2 \over 64 \pi}
M_{\Delta} \sqrt{1 - {4 M_N^2 \over M_\Delta^2}}~.
\label{width}
\end{equation}

Since in our scheme, lepton asymmetry is initially created as an asymmetry in $\tilde{N}$ versus
$\tilde{N}^*$, we are interested in range of model parameters where these decays are essentially
out-of-equilibrium at temperatures around the mass of $\Delta_0$.  For $M_\Delta \sim (10^6-10^8)$ GeV,
this requirement implies that $|f|$ in Eq. (\ref{width}) should obey $|f| \leq 2\cdot (10^{-5}-10^{-4})$.
For such small values of $|f|$, it is important to check if the gauge boson mediated decays of
$\Delta_0$ will have a comparable rate.  To check this, we have computed the total decay width
of $\Delta_0$ scalars into four MSSM fields. These could be four scalars, four fermions, or
two scalars plus two fermions, all of the MSSM.  The total width is given by
\begin{equation}
\Gamma(\Delta_0 \rightarrow \Phi_i^* \Phi_i \Phi_j^* \Phi_j) = {256 \times 4 (g_B/2)^6  \over
360 \times (2 \pi)^5 }\left({M_\Delta^7 \over M_{\cal V_B}^6}\right)~.
\label{width1}
\end{equation}
In Eq. (\ref{width1}), $\Phi_i$ stands for any of the scalar or fermion fields of MSSM.
The factor $256$ arises as $[{\rm Tr} (q_i^2)]^2$, while the factor 4 is to account for the
various types of final states stated above.  We see that these decays are suppressed by
phase space and inverse power of the ${\cal V_B}$ mass.  If we demand that the decays of
$\Delta_0$ given in Eq. (\ref{width}) dominates over the ones in Eq. (\ref{width1}), we arrive
at an inequality $(g_B/2) M_\Delta/M_{\cal V_B} < 1.6 |f|^{1/3}$, or using Eq. (\ref{mass}),
 $|\lambda| \leq 4.5 |f|^{1/3}$.  If $|f| = 10^{-5}$, this translates into a limit
 $|\lambda| \leq 0.1$.  This a rather mild hierarchy, which is quite natural.
We will henceforth assume that the two body decay of $\Delta_0$ into $\tilde{N} \tilde{N}$  dominates
over the four body decay, which would enable us to create lepton asymmetry in $\tilde{N}$.

\section{Spectrum including SUSY breaking}

In the supersymmetric limit we have seen that four real scalar fields belonging to the
$(\Delta_0,~S)$ superfileds are degenerate in mass.  The corresponding fermions are also
degenerate in mass.  This degeneracy will be lifted once SUSY breaking interactions are
taken into account.  One would arrive at two quasi--degenerate Majorana fermions and four
quasi--degenerate real scalar fields.  Their mass splittings and coupling to the $(N,~\tilde{N}$)
fields are crucial for the estimation of the induced lepton asymmetry in $\tilde{N}$.  Here
and in the next section we compute these splittings and couplings.

Soft supersymmetry breaking interactions are introduced in the usual way as in supergravity.  For the
$(\Delta ,~\overline{\Delta},~S,~N)$ sector the relevant soft breaking terms are given by
\begin{eqnarray}
V_{\rm soft} = \{A_\lambda \lambda S \Delta \overline{\Delta} - C_\lambda \lambda M^2S +
{A_ff_{ij} \over 2} \Delta \tilde{N}_i \tilde{N}_j + h.c.\} + m_i^2 \Phi_i^* \Phi_i~.
\label{soft1}
\end{eqnarray}
The dimensional parameters $\{A_\lambda,~A_f,~C_\lambda\}$ will be taken to have values
near the TeV scale.  Mass--splittings within degenerate multiplets will be induced at order
$M_{\rm susy} \sim $ TeV, so we will ignore terms of order $M_{\rm susy}^2$ and higher.  The soft squared
mass parameters $m_i^2$ in Eq. (\ref{soft1}) can then be neglected.

We now minimize the potential including soft SUSY breaking, keeping linear terms in $M_{\rm susy}$.
First we obtain the redefined soft breaking terms after the transformation of Eq. (\ref{subs}) is
applied to Eq. (\ref{soft1}).  This yields
\beq
V_{\rm soft}=\lam M^2(A_{\lam }-C_{\lam })S+
A_{\lam }\lam e^{i\phi_{M^2}}S(\sq{2}|M|\De_0+\fr{1}{2}\De_0^2)+\fr{1}{2}A_ff_{ij}(|M|+\fr{1}{\sq{2}}\De_0)\tl{N}_i\tl{N}_j
+{\rm h.c.}+m_i^2|\Phi_i|^2~.
\la{Vtot}
\eeq
The full potential is given by $V=V_F+V_{\rm Soft}$, with $V_F$ obtained from Eq. (\ref{Wef}) as
\begin{equation}
V_F=|\lam \De_0|^2|\sq{2}|M|+\fr{1}{2}\De_0|^2+|\lam \sq{2}|M|S+\lam S\De_0 +\fr{e^{-i\phi_{M^2}}}{2\sq{2}}f_{ij}\tl{N}_i\tl{N}_j|^2
+|f_{ij}(|M|+\fr{1}{\sq{2}}\De_0)\tl{N}_j|^2~,
\label{VF}
\end{equation}
where we have neglected terms arising from $Y_{\nu }$ coupling.

Minimization of $V$ shows that the field $S$ develops a vacuum expectation value (VEV) of order
$M_{\rm susy}$ given by
\beq
\lan S^*\ran =\fr{C_{\lam }-A_{\lam }}{2\lam^*}e^{i\phi_{M^2}} ~.
\la{sVEV}
\eeq
The shift in the VEV of the $\Delta_0$ field is of order $M_{\rm susy}^2$ and thus negligible.
As a consequence of $\left\langle S \right\rangle \neq 0$, the mass matrix in the fermion
sector spanning $(\tilde{\Delta}_0,~\tilde{S})$ gets modified.  We now have this matrix given by
\begin{equation}
\begin{array}{cc}
 & {\begin{array}{cc}
 &\hs{0.8cm}  \hs{0.8cm} \hs{0.8cm}
\end{array}}\\ \vspace{1mm}
\hs{0.1cm}{\cal M}_{fermi}= e^{i(\phi_{M^2}+\phi_\lambda)}\hs{-0.3cm}
 \begin{array}{c}
\\~  ~ ~
 \end{array}\!\!\!\!\!
\hs{-0.3cm} &{\left(\begin{array}{cc}
|\lambda|\left\langle S \right\rangle & M_\Delta
 \\
M_\Delta & 0
\end{array}\right)}~.
\end{array}  \!\!  ~~~~~
\label{fermi}
\end{equation}
Here we have denoted the phase of $\lambda$ as $\phi_{\lambda}$.  Eq. (\ref{fermi}) leads to two
quasi--degenerate Majorana fermions with masses given by $M_{\psi_{1,2}} = M_\Delta \pm |\lambda \left\langle
S \right\rangle|/2$.

In the bosonic sector, the squared mass matrix spanning
$\l {\rm Re}(\De_0), {\rm Re}(S), {\rm Im}(\De_0), {\rm Im}(S)\r $, is found to be (to order $M_{\rm susy}$)
\begin{equation}
\begin{array}{cccc}
 & {\begin{array}{cccc}
 &\hs{0.8cm} & \hs{0.8cm}& \hs{0.8cm}
\end{array}}\\ \vspace{1mm}
\hs{0.1cm}{\cal M}^2_{boson}= M_\Delta^2\hs{-0.3cm}
 \begin{array}{c}
\\~  ~ \\~  ~\\ ~
 \end{array}\!\!\!\!\!
\hs{-0.3cm} &{\left(\begin{array}{cccc}

 1 &  \kappa_R+\kappa'_{R}&0& \kappa_I-\kappa'_{I}
\\
 \kappa_R+\kappa'_{R}&1&-\kappa'_{I} &0
 \\
 0&-\kappa'_{I} & 1 &-\kappa'_{R}
 \\
 \kappa_I-\kappa'_{I}& 0& -\kappa'_{R} &1
\end{array}\right)}~,
\end{array}  \!\!  ~~~~~
\label{MDT}
\end{equation}

\beq
{\rm with}~~~(\kappa_R, \kappa_I, \kappa'_{R}, \kappa'_{I})=\fr{\sq{2}}{|M|}\l {\rm Re}(\left\langle S \right \rangle),~ {\rm Im}(\left\langle S \right \rangle), ~
{\rm Re}(\fr{A_{\lam }e^{i\phi_{M^2}}}{2\lam^*}),~ {\rm Im}(\fr{A_{\lam }e^{i\phi_{M^2}}}{2\lam^*})\r ~.
\la{sc-matrix}
\eeq

The eigenvalues of the matrix in Eq. (\ref{MDT}) are found to be:
\begin{equation}
M_{X_{1,2}}^2=M_{\De }^2\l 1\pm \De_{12}+\De_{14}\r ~,~~~~M_{X_{3,4}}^2=M_{\De }^2\l 1\mp \De_{12}-  \De_{14}\r
\label{eig}
\end{equation}
with the definitions
\beq
\De_{12}=-\fr{|C_{\lam }+A_{\lam }|}{2M_{\De }}~,~~~~~\De_{14}=\fr{|C_{\lam }-A_{\lam }|}{2M_{\De }}~.
\la{sc-masses}
\eeq
Thus $\Delta_{12} = (M_{X_1}^2 - M_{X_2}^2)/(2M_\Delta^2)$ parametrizes the fractional mass splitting in $X_1$
and $X_2$, and similarly $\Delta_{14}$ in $X_1$ and $X_4$.  These two mass splittings will be relevant for
leptogenesis calculation.  We also note the identities $\Delta_{12} = \Delta_{43}$ and $\Delta_{14} = \Delta_{23}$.
There are two other mass splittings which can be obtained in terms of $\Delta_{12}$ and $\Delta_{14}$,
but those two turn out to be not relevant for leptogenesis.

The mass eigenstates $X_i$ are related to the original states as
\begin{eqnarray}
{\rm Re}(\De_0)&=&\fr{c_{\al}}{\sq{2}}(X_1+X_3)+\fr{s_{\al }}{\sq{2}}(X_2+X_4)~,~~~
{\rm Re}(S)=\fr{c_{\bt }}{\sq{2}}(X_1-X_3)+\fr{s_{\bt }}{\sq{2}}(X_2-X_4)~,
\nonumber \\
{\rm Im}(\De_0)&=&-\fr{s_{\al}}{\sq{2}}(X_1+X_3)+\fr{c_{\al }}{\sq{2}}(X_2+X_4)~,~~~
{\rm Im}(S)=-\fr{s_{\bt }}{\sq{2}}(X_1-X_3)+\fr{c_{\bt }}{\sq{2}}(X_2-X_4).
\end{eqnarray}
Here two mixing angles appear which we denote as ($\alpha, \beta$).  We use the notation $c_\alpha = \cos\alpha,~s_\alpha= \sin\alpha$, etc.  These two angles are given by
\beq
\al =\fr{1}{2}\l \pi + {\rm arg}(C_\lambda +A_\lambda)-{\rm arg}(C_\lambda - A_\lambda)\r ~,~~~
\bt =\fr{\pi }{2}+\phi_{M^2} +\phi_{\lam }+\fr{1}{2}\l {\rm arg}(C_\lambda +A_\lambda)+{\rm arg}(C_\lambda - A_\lambda)\r  ~.
\la{sc-vectors}
\eeq
We shall use these results in the next section where we compute the lepton asymmetry
stored in $\tilde{N}$ arising from the decays of these scalar states.

\section{Cosmological lepton asymmetry}

In our scenario, cosmological lepton asymmetry is generated in the out of equilibrium decays
of the $X_i$ scalars into $\tilde{N}$ and $\tilde{N}^*$, the scalar partners of the right--handed
neutrino.  One loop corrections to the decay
induces CP asymmetry, leading to an asymmetry in $\tilde{N}$ versus $\tilde{N}^*$. This induced
asymmetry is converted to usual lepton asymmetry when $\tilde{N}$ and $\tilde{N}^*$ decay into
leptons and a Higgs boson, which subsequently is converted to baryon asymmetry via
electroweak sphaleron processes.

As shown in Sec. 2, the dominant decay of the $X_i$ scalars will be into final states with $\tilde{N}$
scalars and $N$ fermions, with a smallish coupling $\lambda \leq 0.1$ and $|f| \sim 10^{-5}$.  The tree
level decay diagrams are shown in Fig. 1.  The total decay rate for these decays is given in Eq. (\ref{width}).
The decay of $X_i$, which are real scalars, into final states with opposite lepton number ($-2$ and $+2$)
(see Fig. 1 (a) and (b)) raises the possibility that an asymmetry can be produced in $\tilde{N}$ number.
For $M_\Delta = 10^6-10^8$ GeV and $|f| = 10^{-5}-10^{-4}$, the lepton number violating decays of the
$X_i$ fields will be out of equilibrium.  The efficiency factor in the production of $\tilde{N}$ asymmetry will
then be nearly one.

\begin{figure}[!ht]
\begin{center}
\hs{1.7cm}\includegraphics[scale=0.7]{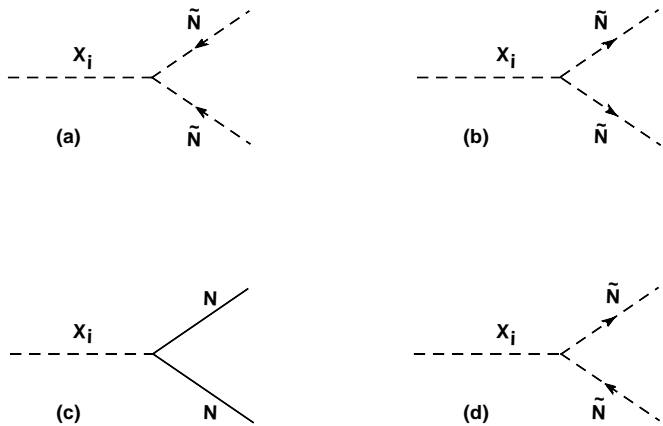}
\vs{0.7cm}
\caption{Tree level decays of $X_i$ scalars into $\tilde{N}, \tilde{N}^*$ and $N$.}
\label{fig:1}
\end{center}
\end{figure}

We now proceed to calculate the induced $\tilde{N}$ asymmetry.  For this purpose we need
to identify the interaction of the $X_i$ fields with $\tilde{N}$.  Since the $X_i$ fields
are quasi--degenerate, the dominant contribution to lepton asymmetry will arise from wave
function corrections shown in Fig. 2.  These corrections have a resonance enhancement, which
is lacking  in the vertex correction diagrams.  SUSY provides the quasi--degeneracy of
$X_i$ fields, which enables us to realize resonant leptogenesis in $\tilde{N}$.  The required
CP violation arises in the model from soft SUSY breaking couplings.  Thus this scenario is
also soft leptogenesis, but with four $X_i$ fields involved in the decay.

From the Lagrangian given in Eqs. (\ref{Vtot}) and (\ref{VF}), one can read off the cubic scalar
interactions relevant for the wave function corrections of Fig. 2.  The couplings of $X_i$ to $\tilde{N}$ is found
to be
\begin{equation}
V^{(3)}=\l \tl{N}\tl{N}F_{\tl{N}\tl{N}i}X_i+{\rm h.c.}\r +|\tl{N}|^2F_{|\tl{N}|i}X_i~,
\label{cubic}
\end{equation}
where we have defined
\begin{eqnarray}
F_{\tl{N}\tl{N}i}&=&\fr{f}{4\sq{2}}\l a_1\!+\!a_2\!+\!M_{\De }e^{i\om },~ -i(a_1\!-\!a_2\!+\!M_{\De }e^{i\om }),~
a_1\!+\!a_2\!-\!M_{\De }e^{i\om },~-i(a_1\!-\!a_2\!-\!M_{\De }e^{i\om })\r_i
\nonumber \\
F_{|\tl{N}|i}&=&\fr{|f||M_N|}{\sq{2}}\l  c_{\al },~s_{\al },~c_{\al },~s_{\al }\r_i
\nonumber \\
{\rm with}~&~&~a_1=\fr{C_{\lam }-A_{\lam}}{2}e^{i\al }~,~~~~a_2=A_fe^{-i\al }~,~~~~\om =\bt -\phi_{\lam }-\phi_{M^2}~.
\la{Fcoupl}
\end{eqnarray}

\begin{figure}[!ht]
\begin{center}
\hs{1cm}\includegraphics[scale=0.9]{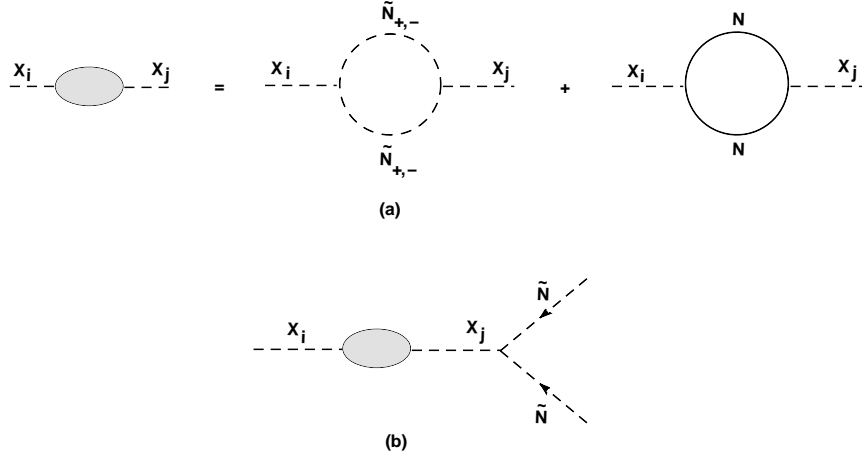}
\vs{0.5cm}
\caption{Loop diagrams generating CP asymmetry in the decay $X_i \rightarrow \tilde{N}^* \tilde{N}^*$.
The blob in (b) corresponds to the resummed two point functions shown in (a).}
\label{fig:2}
\end{center}
\end{figure}

The $\tilde{N}$ and $\tilde{N}^*$ states mix after SUSY breaking.  This splitting effect will show up
in the loops of Fig. 2.  To take these effects into account, we go to the mass eigenbasis of these states $\tilde{N}_{+}$
and $\tilde{N}_{-}$ which are given by
\begin{equation}
\tl{N}_{+}=\fr{1}{\sq{2}}(e^{ix}\tl{N}+e^{-ix}\tl{N}^*)~,~~~~\tl{N}_{-}=\fr{1}{\sq{2}i}(e^{ix}\tl{N}-e^{-ix}\tl{N}^*)~.
\la{Npm-states}
\end{equation}
The phase parameter $x$ in Eq. (\ref{Npm-states}) is defined as
$x=\fr{1}{2}\l \phi_f+{\rm arg}(A_f+\fr{C_{\lam }-A_{\lam }}{2})\r.$
Note that $\tilde{N}_\pm$ are real fields with masses given by
\begin{equation}
M_{\tl{N}_{+}}^2=|M_N|^2+|M_N|\left | A_f+\fr{C_{\lam }-A_{\lam }}{2}\right |~,~~~~
M_{\tl{N}_{-}}^2=|M_N|^2-|M_N|\left | A_f+\fr{C_{\lam }-A_{\lam }}{2}\right |~.
\end{equation}

In the $\tl{N}_a=(\tl{N}_{+}, \tl{N}_{-})$  ($a=\pm $) basis, the cubic scalar interactions can
be written as
\beq
V^{(3)}=(\tl{N}_{+}^2F_{++\hs{0.5mm}i}+\tl{N}_{-}^2F_{--\hs{0.5mm}i}+\tl{N}_{+}\tl{N}_{-}F_{+-\hs{0.5mm}i})X_i~,
\la{tlNNX}
\eeq
where
\begin{eqnarray}
F_{++i}&=&\fr{1}{2}\l e^{-2ix}F_{\tl{N}\tl{N}i}+e^{2ix}F_{\tl{N}\tl{N}i}^*+F_{|\tl{N}|i}\r ~,~~~
F_{--i}=-\fr{1}{2}\l e^{-2ix}F_{\tl{N}\tl{N}i}+e^{2ix}F_{\tl{N}\tl{N}i}^*-F_{|\tl{N}|i}\r ~,
\nonumber \\
F_{+-i}&=&i\l e^{-2ix}F_{\tl{N}\tl{N}i}-e^{2ix}F_{\tl{N}\tl{N}i}^*\r ~.
\la{tlNNXcoupl}
\end{eqnarray}

It is now straightforward to work out the absorptive part of the two point function
arising from diagrams with $\tl{N}$'s in the loops.  We find it to
be
\begin{eqnarray}
\Pi_{ij}^B(p^2)&=&\fr{1}{32\pi }\l 2K_{++}F_{++\hs{0.5mm}i}F_{++\hs{0.5mm}j}+2K_{--}F_{--\hs{0.5mm}i}F_{--\hs{0.5mm}j}
+K_{+-}F_{+-\hs{0.5mm}i}F_{+-\hs{0.5mm}j}\r ~,
\nonumber \\
{\rm where} ~~~~&~&K_{ab}=\l 1-2\fr{M_{\tl{N}_a}^2\!+\!M_{\tl{N}_b}^2}{p^2}+\fr{(M_{\tl{N}_a}^2\!-\!M_{\tl{N}_b}^2)^2}{p^4}\r^{1/2}
\la{PiB}
\end{eqnarray}
When considering $X_i$-decay, one should set $p^2=M_{X_i}^2$.

We will also need the Yukawa couplings of the $X_i$ fields with the $N$ fermions.  It is given by
\begin{eqnarray}
{\cal L}_{NNY}&=&NNY_FX+{\rm h.c.}
\nonumber \\
{\rm with}~~&~&~Y_F=\fr{fe^{-i\al }}{4\sq{2}}\l 1,~i,~1,~i\r ~.
\la{YF}
\end{eqnarray}
The absorptive part arising through the fermionic loop in Fig. 2 is found to be
\beq
\Pi_{ij}^F(p^2)=\fr{1}{16\pi }\sq{1-4\fr{M_N^2}{p^2}}\left [p^2(Y_F^{\dag }Y_F+Y_F^TY_F^*)_{ij}-
2M_N^2(Y_F^TY_Fe^{-2i\phi_f}+Y_F^\dag Y_F^*e^{2i\phi_f})_{ij}\right ]~.
\la{PiF}
\eeq
With these, we have for example, for the absorptive part of $\Pi_{12}$,
\beq
\Pi_{12}=\Pi_{12}^B+\Pi_{12}^F\simeq \fr{|f|^2}{32\pi }\fr{\hat{A_1}}{4M_{\De }}\sq{1-4\fr{M_N^2}{M_{\De }^2}}M_{\De }^2~,
\la{PiB12}
\eeq
where $\hat{A_1}$ is defined in Eq. (\ref{def-A-hats}).

We now combine these results to compute $\epsilon_{\tilde{N}}$, the $\tilde{N}$ asymmetry parameter
defined as
\begin{equation}
\epsilon_{\tilde{N}} = \sum_{i}{\Gamma(X_i \rightarrow \tilde{N}\tilde{N}) - \Gamma(X_i \rightarrow \tilde{N}^*
\tilde{N}^*) \over \Gamma(X_i \rightarrow \tilde{N}\tilde{N}) + \Gamma(X_i \rightarrow \tilde{N}^*
\tilde{N}^*)}~.
\end{equation}
We find it to be
\beq
\epsilon_{\tilde{N}} =4\left [ \fr{2\De_{12}\Ga/M_{\De }}{4\De_{12}^2+(\fr{\Ga }{2M_{\De }})^2}\cdot \fr{\hat{A}_1}{M_{\De }}+
\fr{2\De_{14}\Ga/M_{\De }}{4\De_{14}^2+(\fr{\Ga}{2M_{\De }})^2}\cdot \fr{\hat{A}_2}{M_{\De }}\right ]~,
\la{CPasym}
\eeq
where $\Ga $ is a total decay width [i.e. $\Ga (X_i\to {\rm everything})$].
Here we have defined two effective $A$--parameters as follows:
\begin{eqnarray}
\hat{A}_1&=&|A_f|\sin \phi_1-2\left |A_f+\fr{C_{\lam }-A_{\lam }}{2}\right |\fr{(M_N/M_{\De })^2}{1-4(M_N/M_{\De })^2}\sin \phi_2
\nonumber \\
\hat{A}_2&=&-2\left |A_f+\fr{C_{\lam }-A_{\lam }}{2}\right |\fr{(M_N/M_{\De })^2}{1-4(M_N/M_{\De })^2}\sin \phi_3
\la{def-A-hats}
\end{eqnarray}
The phases appearing in Eq. (\ref{CPasym}) are related to the original phases in the model through the relations
\vs{-0.4cm}
$$
\phi_1={\rm arg}(A_f)-{\rm arg}(C_\lambda +A_\lambda)~,~~~
\phi_2={\rm arg}(A_f + {C_\lambda - A_\lambda \over 2})-{\rm arg}(C_\lambda +A_\lambda)~,
$$
\beq
\phi_3={\rm arg}(A_f + {C_\lambda - A_\lambda \over 2})-{\rm arg}(C_\lambda - A_\lambda) ~.
\la{entries}
\eeq

It should be mentioned that the asymmetry given in Eq. (\ref{CPasym}) includes fermionic
and bosonic loop contributions.  It turns out that the fermionic loop is entirely canceled by
the bosonic loop, the left-over piece from the bosonic loop is what is given in Eq. (\ref{CPasym}).
This cancelation is not surprising, since the fermion loop corrections do not feel the effects
of SUSY breaking.  We also note that the off--diagonal $\Pi_{ij}$ have one power of $M_{\rm susy}/M_X$
suppression, so the decay vertex has to be supersymmetric.  This feature simplifies the calculations
somewhat.  In Eq. (\ref{CPasym}) we have added the asymmetry arising from all four of the $X_i$ scalar
fields.

In principle, the decays of the Higgsinos $(\tilde{\Delta}_0, \tilde{S})$ into $\tilde{N}$ and $N$
can create an asymmetry in $\tilde{N}$.  However, we find that there is not sufficient CP violation
in these decays in the minimal model.

Now we are ready to estimate the lepton asymmetry created by $\tl{N}$-decays at the second stage
where $\tilde{N}$ decays into a lepton and a Higgs boson.
Note that lepton asymmetry between $\tl{N}$ and $\tl{N}^*$ will be completely converted into lepton asymmetry in the MSSM sector. There is however one peculiarity related to SUSY.
$\tl{N}$ has two primary decay channels $\tl{N}\to L\tl{H}_u$ and $\tl{N}\to \tl{L}^*H_u^*$.
Since the rates of these processes are the same due to SUSY  (at zero temperature),
the lepton asymmetries created from these decays cancel each other.
However, with  $T\neq 0$ the cancelation is only partial (due to temperature effects which
explicitly break SUSY) and one has
\beq
\tl{\ep }=\ep (\tl{N}\to L\tl{H}_u)\De_{BF}~,
\la{ep-non-zero}
\eeq
with the temperature dependent factor $\De_{BF}$ given in Ref. \cite{giudice}.
Now, the baryon asymmetry created from the lepton asymmetry due to $\tl{N}$ decays is:
\beq
\fr{\tl{n}_B}{s}\simeq -8.6\cdot 10^{-4}\fr{\tl{\ep }}{\De_{BF}}\eta =
-8.6\cdot 10^{-4}\epsilon_{\tilde{N}}\eta ~,
\la{sc-asym}
\eeq
where we have taken into account  an effective number of degrees of freedom, including one RHN superfield, to be
$g_*=225$. In the last stage of Eq. (\ref{sc-asym}) we have substituted  $\tl{\ep }$ by  $\epsilon_{\tilde{N}}$ - the $\tl{N}$ asymmetry created at the first stage by $X_i$-decays.
$\eta $ is an efficiency factor which depends on $\tl{m}\simeq \fr{v_u^2}{M}Y_{\nu }^2$, and  which
takes into account temperature effects by integrating the Boltzmann equations \cite{giudice}. For instance, efficiency $\eta $ reaches its maximal value, $\eta \approx 0.1$  for $\tl{m}\approx 10^{-3}$~eV. Thus, in order to generate the experimentally observed asymmetry  $\l \fr{n_B}{s}\r_{\rm exp}=(8.75\pm 0.23)\cdot 10^{-11}$,
we need to have $\epsilon_{\tilde{N}}\stackrel{>}{_\sim }10^{-6}$.
Going back to Eq. (\ref{CPasym}), we see that an enhancement of $\epsilon_{\tilde{N}}$ will happen for small values of $\De_{ij}$. The natural values of these parameters are $\sim M_{susy}/M_{\De }$. However, some cancelation can make either of these parameters
smaller.  Assuming that this happens for $\De_{12}$, with the parametrization $\De_{12}= \de_{12} M_{susy}/M_{\De }$ and $\hat{A}_1=\de_1M_{susy}$
we have $\epsilon_{\tilde{N}}\approx 2\de_1\Ga/(\de_{12} M_{\De })$. On the other hand, out of equilibrium decay of $X_i$ states requires $\Ga \stackrel{<}{_\sim }H=1.7\sq{g_*}M_{\De }^2/M_{\rm Pl}$. Therefore, we have
$\epsilon_{\tilde{N}}\stackrel{<}{_\sim }3.4\sq{g_*}\de_1M_{\De }/(\de_{12} M_{\rm Pl})$. With the choice $\de_1\approx 3$ and
$\de_{12} \approx 1/300$
and $M_{\De }\simeq 10^{8}$~GeV, we obtain $\epsilon_{\tilde{N}}\simeq 10^{-6}$. This has been achieved by the suppressed value
$\de_{12}$, which does not seem to be natural.  Similar situation occurs in the soft leptogenesis scenario. However, note that within our setup we do not need to constrain the value of the Dirac Yukawa coupling $Y_{\nu }$ very much. The only real constraint is that 
 $\tl{N}$ decays out of equilibrium,  which requires $\Ga \stackrel{<}{_\sim }H$. 

We conclude with a few remarks.  We have kept corrections linear in $M_{\rm susy}/M_\Delta$ in the computation
of CP asymmetry, and not any higher
powers.  It is known that if the mass of the decaying field is close to the SUSY scale, second order vertex
corrections can be important proportional to the mass of the MSSM gaugino \cite{concha}.  In our scheme, these vertex corrections do not exist, since the $B-L$ gaugino has decoupled and since $\tilde{N}$ does not couple to
MSSM gauginos.  A natural question to ask is whether the soft SUSY breaking corrections that induce lepton
asymmetry can also lead to excessive CP violation in electron and neutron dipole moments.  With universal
soft breaking mass parameters there is a potential problem.  We note that if the theory is embedded in
SUSY left--right model, then all the Dirac Yukawa couplings and $A$--terms are hemitian due to parity
symmetry.  That will make all EDM contributions vanishingly small \cite{dutta}. On the other hand, parity symmetry
implies that the Majorana--type couplings (such as $A_f$ and $f$ in our model) are complex symmetric, which
can serve to induce the lepton asymmetry.

\section*{\small Acknowledgement}
\vs{-0.2cm}  The work is supported in part by DOE grant
DE-FG02-04ER41306 and DE-FG02-ER46140. Z.T. is also partially
supported by GNSF grant 07\_462\_4-270.

\end{document}